%% file: main.tex
\documentclass[conference]{IEEEtran}
\IEEEoverridecommandlockouts

\usepackage{cite}
\usepackage{amsmath,amssymb,amsfonts}
\usepackage{algorithmic}
\usepackage{graphicx}
\usepackage{subcaption}
\usepackage{textcomp}
\usepackage{xcolor}
\usepackage{hyperref}

\usepackage{booktabs}
\usepackage{multirow}
\usepackage{enumitem}

\newcommand{\ie}{\textit{i.e.}, } 
\newcommand{\eg}{\textit{e.g.}, } 

\newcommand{\NA}{---} 

\newcommand\blfootnote[1]{%
  \begingroup
  \renewcommand\thefootnote{}%
  \footnotetext{#1}%
  \endgroup
}

\newcommand{\expnumber}[2]{{#1}\mathrm{e}{#2}}

\def\BibTeX{{\rm B\kern-.05em{\sc i\kern-.025em b}\kern-.08em
    T\kern-.1667em\lower.7ex\hbox{E}\kern-.125emX}}

\begin{document}

\title{Empowering Compact LLMs with Fusion of Layer-wise Exits for Recommendation}


\author
{
 Xurong Liang{\small$^\dag$}\hspace*{10pt}
 Tong Chen{\small$^\dag$}\hspace*{10pt}
 Quoc Viet Hung Nguyen{\small$^\perp$}\hspace*{10pt}
 Jianxin Li{\small$^\S$}\hspace*{10pt}
 Xiangliang Zhang{\small$^\ddagger$}\hspace*{10pt}
 Hongzhi Yin{\small$^{\dag *}$}\\
 
 \fontsize{10}{10}\selectfont\itshape
 $^\dag$The University of Queensland, Australia,
 {\fontsize{9}{9}\selectfont\ttfamily\upshape
 \{xurong.liang,tong.chen,h.yin1\}@uq.edu.au}\\
 
 \fontsize{10}{10}\selectfont\itshape
 $^\perp$Griffith University, Australia,
 {\fontsize{9}{9}\selectfont\ttfamily\upshape
 henry.nguyen@griffith.edu.au}\\
 
\fontsize{10}{10}\selectfont\itshape $^\S$Edith Cowan University, Australia,
{\fontsize{9}{9}\selectfont\ttfamily\upshape jianxin.li@ecu.edu.au}\\
 
 \fontsize{10}{10}\selectfont\itshape
 $^\ddagger$University of Notre Dame, USA,
 {\fontsize{9}{9}\selectfont\ttfamily\upshape
 xzhang33@nd.edu}
}

\maketitle

\blfootnote{*Hongzhi Yin is the corresponding author.}

\begin{abstract}
\input{sections/abstract}
\end{abstract}


\section{Introduction}
\input{sections/intro}

\section{Related Work}
\input{sections/related}

\section{Method}
\input{sections/method}

\section{Experiment}
\input{sections/experiment}

\section{Conclusion}
\input{sections/conclusion}

\section*{Acknowledgments}
This work is partially supported by the Australian Research Council (Grant No. DE230101033, FT210100624, LP230200892, LP240200546, LP250200778, DP260100326 and DP240101814).

\bibliographystyle{IEEEtran}
\bibliography{custom}

\end{document}

%% file: sections/abstract.tex
Large language model-based recommender systems (LLM-RSs) have demonstrated remarkable capabilities, but are also known to be computationally unsustainable for real-world applications. 
While utilizing compact LLMs (\eg the ones with 3 billion or fewer parameters) is a direct workaround, 
the reduced model capacity often necessitates
methods like reasoning and knowledge distillation that exacerbate inference latency or rely on larger LLM counterparts.
Combined with their autoregressive nature, those generative LLM-RSs inevitably face a severe scalability bottleneck. In contrast, discriminative LLM-RSs use LLMs as embedding generators to support efficient full-corpus item ranking through embedding similarity, but are prone to the limited expressiveness of compact LLMs and lack structural adaptivity when extracting patterns from multifaceted user interactions. In this paper, we propose the Fusion of Layer-wise Exits for Sequential Recommendation (FLEXRec), a discriminative LLM-RS paradigm that 
supports scalable full-corpus item ranking
and maintains high effectiveness when a compact LLM is in use. 
Specifically, treating each transformer block in an LLM as a \textit{layer}, we bypass the single-output design of existing LLM-RSs by inserting prediction heads (\ie \textit{exits}) at each layer, allowing for an ensemble of layer-wise exits to capture a mixture of user preference patterns at different depths and granularities. 
We further design an adaptive continuous router (AC-Router), which enables the model to dynamically select both the number and identity of exits conditioned on each input user sequence.
The AC-Router is regularized by a novel target-$k$ hinge loss to enforce routing sparsity within a designated range of activated exits. Extensive experiments on three real-world datasets utilizing compact LLM backbones (\ie Qwen 3 1.7B and Llama 3.2 3B) demonstrate that FLEXRec achieves state-of-the-art accuracy among competing methods while remaining highly efficient. The codebase of FLEXRec is available at \url{https://github.com/xurong-liang/FLEXRec}.

%% file: sections/intro.tex
Large language model-based recommender systems (LLM-RSs) have attracted surging interest from both academia \cite{bao2023tallrec, zhang2025collm} and industry \cite{yu2024cosmo, liang2025context} due to the mutual benefit between textual context modeling and user preference learning. To take full advantage of language modeling capabilities, most existing LLM-RSs rely on a sizable LLM backbone (\eg 7 or 13 billion parameters \cite{zhang2025collm, liao2024llara}). However, resorting to heavily parameterized LLMs to handle complex user interaction data is computationally unsustainable and difficult to scale \cite{kaplan2020scaling}. 
Naturally, this has stimulated the trend of using LLMs with a more compact  architecture (\eg 1.7 billion parameters \cite{yang2025qwen3}) as the recommendation backbone. Nevertheless, as LLMs' semantic comprehension capabilities are directly linked to their scales  \cite{kaplan2020scaling}, 
the resulting LLM-RSs are subject to weakened ability to capture sophisticated user behavioral patterns.

To counteract the limited modeling capacity inherent to compact LLMs, recent literature has explored two primary paradigms: reasoning and knowledge distillation. On the one hand, reasoning-enhanced frameworks introduce complicated reasoning techniques, such as chain-of-thought \cite{yue2025cot4rec} or generative behavioral comprehension \cite{kong2026think}, to force the compact model to generate intermediate rationales or context to augment the recommendation. On the other hand, knowledge distillation frameworks \cite{cui2024distillation, kim2025lost} train a lightweight student model that mimics the behaviors of a more capable (and commonly larger) teacher model. Although both can improve the recommendation accuracy with a compact LLM, they face severe operational limitations. Knowledge distillation requires expensive offline training and incurs high costs for hosting a powerful teacher backbone for knowledge transfer, while complex reasoning paradigms introduce substantial computational steps that severely exacerbate inference latency. In addition, 
as they mostly fall into the \textit{generative} recommendation paradigm \cite{wu2024survey, kim2025lost}, the compounding latency of autoregressive text generation is unavoidable. Together, those factors render such methods difficult to scale to full-corpus item ranking, which is a critical task in real-world systems where the model must retrieve the top-$N$ items out of the entire dataset for each user. Even when frameworks generate engineered semantic token sequences rather than natural text \cite{wang2024learnable}, they rely on constrained sequence decoding mechanisms (\eg Trie-based search \cite{de2020autoregressive}) to ensure valid item identifier combinations, which drastically caps system throughput compared with the traditional recommender systems.

To resolve the scalability bottleneck and facilitate full-corpus item ranking, a different branch of LLM-RSs adopts a \textit{discriminative}, embedding-based recommendation paradigm \cite{li2023e4srec, kim2025lost, bao2025heterogeneous}. In this paradigm, users and items are projected into the LLM semantic space solely for interaction modeling, allowing next-item predictions to be generated by computing affinity scores directly on the LLM embeddings. Since this strategy completely bypasses autoregressive text decoding, real-time ranking on the full item catalog becomes computationally viable. E4SRec \cite{li2023e4srec} is an iconic framework in this category, capturing user preference embeddings from interaction records with an LLM
to compute user-item affinity scores. Despite the efficiency benefits of this discriminative LLM-RS paradigm, it suffers from two key limitations when a compact LLM backbone is in use. First, as per the scaling law \cite{kaplan2020scaling, hong2024scale}, the quality of the contextualized embeddings degrades as model parameters shrink, impeding recommendation accuracy. Second, unlike generative methods that can flexibly steer the way LLM reasons with text tokens (\eg adaptive reasoning), in the discriminative paradigm, every user sequence is embedded and processed through exactly the same architectural pipeline 
regardless of how simple or complex the underlying preference pattern is, severely hindering model generalizability across diverse user populations.

To this end, we seek to retain the efficiency of compact LLM backbones in discriminative LLM-RSs without compromising effectiveness.
In this paper, we propose the \textbf{F}usion of \textbf{L}ayer-wise \textbf{Ex}its for Sequential \textbf{Rec}ommendation (FLEXRec). Adopting the discriminative paradigm for full-corpus item ranking capacity, FLEXRec breaks away from the single-output constraint implemented by classic discriminative models \cite{li2023e4srec}. In existing discriminative LLM-RSs, a user's embeddings are propagated through all transformer layers\footnote{Throughout this paper, for simplicity, we refer to each individual transformer block as a \textbf{layer}.} to form the final prediction. Instead of discarding these intermediate representations, FLEXRec explicitly extracts the user preference embeddings generated at \textit{every} layer of the LLM backbone. This architecture enables a prediction head to be inserted after each layer, treating each depth as an independent \textit{exit} point for user-item affinity score prediction. One can then combine the prediction score distributions across multiple exits to generate a fused score distribution. This layer-wise ensemble strategy is inspired by the observation that different architectural depths in an LLM capture distinct, hierarchical semantic patterns \cite{skean2025layer, gromovun2025reasonable}. In the context of sequential recommendation, this hierarchical capacity aligns with the need to model both immediate, short-term item transitions and high-level, long-term user intents \cite{kang2018self, sun2019bert4rec}. By aggregating score distributions generated from a diverse set of representations, FLEXRec introduces rich, complementary context that effectively compensates for the limited parameter breadth of small model sizes.

Furthermore, to enable our model to adaptively reconfigure its structure for capturing each user's nuanced preference patterns, we design an adaptive continuous router (AC-Router) to control the discrete, input-adaptive fusion of these layer-wise exits. Acting as a dynamic gating network inspired by mixture-of-experts (MoE) architectures \cite{mu2025comprehensive}, 
the AC-Router conditions its routing decision on the incoming user sequence to determine
not only \textit{which} exits are activated, but exactly \textit{how many} are needed. 
For simpler interaction patterns, the router can activate fewer exits, whereas more intricate patterns can be modeled using richer exit combinations.
To prevent redundant fusion and enforce computational sparsity, we introduce a joint optimization framework guided by a novel target-$k$ hinge loss that softly penalizes the router when active exit counts fall outside a designated range. As such, our innovative design allows FLEXRec to achieve unparalleled recommendation effectiveness without compromising inference throughput.

We summarize our main contributions below:
\begin{itemize}
    \item We innovatively propose transitioning from static, single-exit LLM recommenders to a dynamic, multi-layer fusion paradigm. By adaptively ensembling prediction score distributions from various LLM propagation depths, we effectively compensate for the limited language modeling capabilities of compact LLM backbones, facilitating highly accurate and scalable sequential recommendation for diverse user interaction behaviors.
    \item We design the FLEXRec framework equipped with an adaptive continuous router (AC-Router) that utilizes continuous ReLU routing to dynamically assign variable computational depth on a per-sequence basis. To ensure robust end-to-end training, we introduce a joint optimization framework, including a novel target-$k$ hinge loss, to gracefully govern the exit fusion behavior with no additional parameter overhead.
    \item We conduct extensive experiments on three real-world datasets to compare the performance of FLEXRec against state-of-the-art traditional and LLM-based sequential recommenders. Utilizing small-scale backbones (\ie Qwen 3 1.7B and Llama 3.2 3B), the results indicate that FLEXRec achieves superior recommendation accuracy and diverse layer utilization without compromising inference efficiency.
\end{itemize}

%% file: sections/related.tex
In this section, we outline relevant work across two main areas, namely LLM-based recommendation and MoE.

\subsection{LLM-Based Recommendation}
Existing LLM-RSs fall into two paradigms \cite{wu2024survey}: generative models that autoregressively decode item texts or tokens \cite{liao2024llara, kim2024large, zhang2025collm, bao2025bi, tan2024idgenrec, wang2024learnable}, and discriminative models \cite{li2023e4srec, kim2025lost} that project items into the LLM space to compute affinity scores. While highly effective, these methods typically rely on massive backbones (\eg 7 billion parameters), making them computationally unsustainable for handling diverse, real-world user interactions at scale.

To address this, lightweight approaches employ model compression (\eg pruning \cite{sun2024simple, ma2023llm} and quantization \cite{xiao2023smoothquant, dettmers2022gpt3}) or knowledge distillation \cite{cui2024distillation, kim2025lost} to condense heavy models into more compact architectures. 
Complementary research on lightweight recommendation has explored compositional embedding, embedding pruning and lightweight graph-based embeddings to reduce the storage and computational costs of conventional recommenders \cite{liang2023learning, tran2025device, qu2026sparse, liang2024lightweight, liang2025lightweight}.
Alternatively, some work retains compact LLMs but introduces complex reasoning paradigms to inject supplementary context \cite{yue2025cot4rec, kong2026think}.

Crucially, these methods enforce a static architectural capacity, extracting ranking signals from a single, fixed output layer. In contrast, FLEXRec dynamically fuses exits from multiple propagation depths on a per-sequence basis. This multi-layer fusion strategy effectively compensates for the limited capacity of compact LLM backbones, significantly enhancing recommendation performance without relying on costly reasoning steps.

\subsection{Mixture-of-Experts}
Mixture-of-Experts (MoE) architectures scale model capacity without proportionally increasing computational cost by routing inputs to specialized sub-models (experts) \cite{cai2025survey, dai2024deepseekmoe, mu2025comprehensive}. While Top-$k$ routing \cite{shazeer2017outrageously, fedus2022switch, dai2024deepseekmoe, zoph2022st} remains the dominant strategy, it strictly allocates a fixed number of experts per instance.

To overcome this rigidity, dynamic routing frameworks allocate a variable number of experts \cite{huang2024harder, guo2025dynamic}. Methods like continuous routing \cite{wang2025remoe, zhuang2026ld} utilize fully differentiable functions, effectively mitigating the optimization difficulties of discrete sorting operations. While traditional MoE research focuses on scaling internal network width, FLEXRec uniquely adapts continuous MoE routing to facilitate the fusion of intermediate layer-wise exits, transforming the MoE paradigm into an adaptive depth-pruning and representation-fusion engine.

%% file: sections/method.tex
In this section, we present the design details of our proposed framework FLEXRec.

\subsection{Problem Formulation}
We consider a commonly used sequential recommendation setup. Let $\mathcal{U}$ and $\mathcal{I}$ respectively denote the sets of users and items. For an arbitrary user $u \in \mathcal{U}$, her historical item interaction sequence, sorted in chronological order, is denoted as $\mathbf{s}^u = (s_1^u, s_2^u, \dots, s_t^u)$, where $s_k^u \in \mathcal{I}$ represents the $k$-th item interacted with by user $u$, and $t$ is the length of the sequence. The objective of a sequential recommender system is to estimate the likelihood of interacting with each item $i \in \mathcal{I}$, so as to form a top-$N$ ranking list at time $t+1$, given her historical sequence $\mathbf{s}^u$.

\subsection{Backbone Recommender System} \label{sec:backbone}
We follow \cite{li2023e4srec} to build a base LLM encoder for discriminative recommendation. Specifically, FLEXRec first pretrains a classic ID-based sequential recommender, namely SASRec \cite{kang2018self}, to obtain collaborative item embeddings for all items in the dataset, denoted as $\mathbf{E}_{pt} \in \mathbb{R}^{|\mathcal{I}| \times d_{pt}}$, where $d_{pt}$ is the embedding dimension of the SASRec model. 

To map these embeddings into the semantic language space of the LLM, a single-layer MLP linear projection $f: \mathbb{R}^{d_{pt}} \mapsto \mathbb{R}^{d_{LLM}}$ is introduced for item ID injection, where $d_{LLM}$ is the hidden size of the LLM backbone. Given a user's sequence $\mathbf{s}^u$, the corresponding sequence of ID-injected item embeddings is gathered as $\mathbf{E}_{inj} = [\mathbf{e}_{s_1^u}, \dots, \mathbf{e}_{s_t^u}] \in \mathbb{R}^{t \times d_{LLM}}$. 

For language modeling, E4SRec constructs the input to the LLM backbone by concatenating the embeddings of the task instructions, the injected item IDs, and a final response token:
\begin{equation}
    \mathbf{E}^0 = 
    \begin{bmatrix}
        \mathbf{E}_{inst}\\ \mathbf{E}_{inj} \\ \mathbf{E}_{resp}
    \end{bmatrix},
\end{equation}
where $\mathbf{E}_{inst} \in \mathbb{R}^{l_{inst} \times d_{LLM}}$ and $\mathbf{E}_{resp} \in \mathbb{R}^{1 \times d_{LLM}}$ are the initial token embeddings of the instruction text (with length $l_{inst}$) and the response token, respectively. These concatenated embeddings $\mathbf{E}^0$ act as the input to the first layer of the LLM.

After that, $\mathbf{E}^0$ are propagated through the $L$ layers of the LLM backbone, generating the contextualized embeddings $\mathbf{E}^l$ after each layer $l$. In the vanilla E4SRec architecture, only the contextualized embeddings from the final layer $\mathbf{E}^{L}$ are utilized. A single-layer MLP prediction head $h_{L}: \mathbb{R}^{d_{LLM}} \mapsto \mathbb{R}^{|\mathcal{I}|}$ takes the propagated representation of the response token to compute raw affinity scores over all items in the dataset. A Softmax function is then applied to these scores to yield the predicted probability distribution $\mathbf{O}^{L} \in \mathbb{R}^{|\mathcal{I}|}$:
\begin{equation} \label{eq:E4SRec_score_gen}
        \mathbf{O}^{L} = \text{Softmax}(h_{L}(\mathbf{e}^L)), \quad\mathbf{e}^L = \mathbf{E}^L_{[-1]},
\end{equation}
where $\mathbf{E}^L_{[-1]}$ denotes the extraction of the final row vector (corresponding to the response token) from the layer's output sequence, yielding the extracted vector $\mathbf{e}^L \in \mathbb{R}^{d_{LLM}}$. 

\textbf{Multi-Layer Extension.} As established in our motivation, relying exclusively on the final layer of a compact LLM backbone for score distribution generation underutilizes the model's limited capacity. Because different architectural depths encapsulate distinct hierarchical processing capabilities, ranging from localized short-term item transitions in shallow layers to abstract, long-term user intents in deeper blocks, extracting representations at every depth provides rich, complementary sequential context. To harness this diverse ranking logic, we extend the E4SRec architecture by inserting a series of prediction heads $\mathcal{H}=\{h_1, \dots, h_L\}$, effectively creating multiple \textit{exits} along the network's forward pass. Specifically, $h_l$ is the prediction head (\ie exit) inserted immediately after the $l$-th layer of the LLM backbone. Generalizing Equation \ref{eq:E4SRec_score_gen}, the probability distribution $\mathbf{O}^l$ generated at exit $l$ is computed as:
\begin{equation} \label{eq:score_gen}
    \mathbf{O}^l = \text{Softmax}(h_l(\mathbf{e}^l)), \quad \mathbf{e}^l = \mathbf{E}^l_{[-1]}, \quad \text{for } l = 1, 2, \dots, L.
\end{equation}
This set of exits $\mathcal{H}$ subsequently serves as the candidate pool for our adaptive routing and fusion mechanism.

\subsection{Adaptive Continuous Router} \label{sec:ACR}
To enable the dynamic fusion of layer-wise exits for sequential recommendation, we introduce the adaptive continuous router (AC-Router). As outlined in Section \ref{sec:backbone}, our backbone LLM is augmented with a diversified set of exits within its intermediate layers. Conditioned on each user sequence, AC-Router dynamically selects an optimal subset of LLM layers for preference modeling. 
For each historical sequence $\mathbf{s}^u$, the AC-Router takes the response token embedding generated immediately after the first LLM layer, denoted as $\mathbf{e}^1 = \mathbf{E}^1_{[-1]}$, to compute raw routing logits $\mathbf{z} \in \mathbb{R}^L$:
\begin{equation}
    \mathbf{z} = \mathbf{e}^1 \mathbf{W}_z,
\end{equation}
where $\mathbf{W}_z \in \mathbb{R}^{d_{LLM} \times L}$ is a learnable projection matrix that maps the hidden dimension to the number of available layer-wise exits $L$.

In traditional Mixture-of-Experts (MoE) architectures, Top-$k$ routing is the dominant expert selection strategy due to its simplicity \cite{shazeer2017outrageously, fedus2022switch, dai2024deepseekmoe}. However, Top-$k$ routing forces every input sequence to fuse exactly $k$ distributions \cite{zhuang2026ld, guo2025dynamic, wang2025remoe}, which imposes unnecessary computational overhead on simpler sequences yet artificially limits the diversity of layer-wise exits when tackling complex ones. Furthermore, without proper workarounds, its non-differentiable nature is known to interfere with end-to-end training \cite{wang2025remoe}.

As such, our proposed AC-Router utilizes a piecewise activation function \cite{wang2025remoe} to generate sparse routing weights. Specifically, we apply the Rectified Linear Unit (ReLU) function over the raw logits at the vector level, \ie $\mathbf{w} = \text{ReLU}(\mathbf{z})$. In the context of MoE solutions, this continuous routing paradigm increases the fusion diversity as it controls not only the identity but also the number of experts (\ie exits in our case) used for each input sample. While Top-$k$ routing restricts a sequence to $\binom{L}{k}$ possible exit fusion combinations, applying ReLU exponentially expands the combinatorial space to $2^L - 1$. This expanded capacity allows the AC-Router to flexibly tailor the semantic depth and ensemble composition to the specific 
characteristics of each user sequence.

However, the ReLU activation introduces a vulnerability: if all raw logits in $\mathbf{z}$ fall below zero, the routing weights $\mathbf{w}$ will collapse entirely, resulting in no exits being selected. To guarantee computational stability, we apply a conditional shift constraint to the raw logits prior to activation to compute the routing weights $\mathbf{w}$:
\begin{equation} \label{eq:safe_logits}
    \begin{aligned}
        s &= \min(\max(\mathbf{z}) - \epsilon, 0),\\
        \hat{\mathbf{z}} &= \mathbf{z} - \text{sg}(s),\\
        \mathbf{w} &= \text{ReLU}(\hat{\mathbf{z}}),
    \end{aligned}
\end{equation}
where $\epsilon$ is a precision safety margin (\eg $\epsilon = 0.1$) and $\text{sg}(\cdot)$ denotes the stop-gradient operator, which prevents the optimizer from artificially inflating the scalar shift $s$ during backpropagation. This constraint guarantees that the maximum logit always survives with at least magnitude $\epsilon$. 

We apply normalization to the gating weights for score fusion. The final gating weights are defined as $\mathbf{g} = \{g_l\}_{l=1}^{L}$, such that:
\begin{equation} \label{eq:gating_weight}
    g_l = \frac{w_l}{\sum^L_{m=1} w_m + \xi},
\end{equation}
where $w_l \in \mathbf{w}$ is the routing weight for the $l$-th layer, and $\xi$ is a small constant (\eg $\xi = \expnumber{1}{-9}$) to prevent zero-division. For any $g_l \in \mathbf{g}$, if $g_l = 0$, then the corresponding exit $h_l$ is disabled for the current sequence.

Based on the sparse gating weights $\mathbf{g}$, the set of active exits for a given user sequence is defined as:
\begin{equation} \label{eq:active_layers}
    \mathcal{A} = \{l \mid g_l > 0\}.
\end{equation}
In FLEXRec, the maximum required propagation depth is determined by the deepest active layer, $l_{max} = \max(\mathcal{A})$. The LLM sequentially propagates embeddings only up to $l_{max}$ and all subsequent layers and exits are skipped. 

Finally, the model fuses the predicted score distributions from the activated exits using their normalized gating weights:
\begin{equation} \label{eq:fused_distribution}
    \mathbf{O}^{fused} = \sum_{l \in \mathcal{A}} g_l \mathbf{O}^l.
\end{equation}
The fused score distribution $\mathbf{O}^{fused}$ will be used to produce the ultimate Top-$N$ ranking list for recommendation.

\begin{figure*}[t]
    \centering
    \includegraphics[width=.8\textwidth, height=.5\textheight, keepaspectratio]{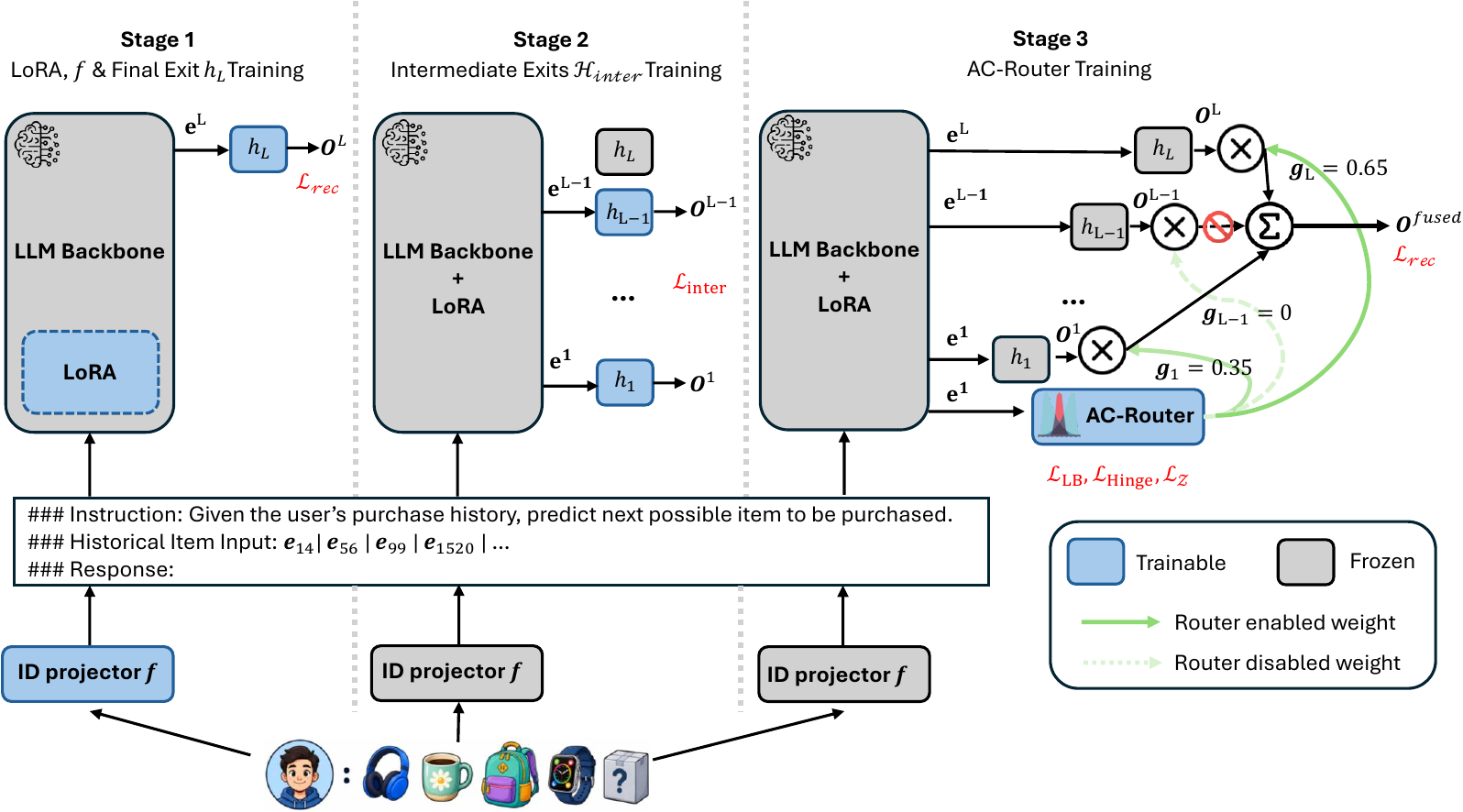}
    \caption{Overview of the FLEXRec framework and its three-stage training. Stage 1 trains the base components via $\mathcal{L}_{rec}$. Stage 2 warms up intermediate exits $\mathcal{H}_{inter}$ via $\mathcal{L}_{inter}$. Stage 3 optimizes the AC-Router using $\mathcal{L}_{rec}$ alongside router-specific loss terms ($\mathcal{L}_{LB}$, $\mathcal{L}_{Hinge}$, $\mathcal{L}_Z$). In the given sequence example, solid green arrows depict active routing weights assigned by the AC-Router, whereas dotted arrows indicate unselected exits that do not contribute to the final fused distribution $\mathbf{O}^{fused}$.}
    \label{fig:framework}
    \vspace*{-3mm}
\end{figure*}

\subsection{Learning to Fuse Layer-wise Exits} \label{sec:ACR_learn}
While the AC-Router establishes the forward-pass mechanics for dynamic layer fusion, it requires carefully formulated loss objectives to correctly map user sequence features to optimal exit combinations. We introduce the following three distinct loss terms to shape the router's behavior:

\textbf{Target-$k$ Hinge Loss.} 
This is our key innovation that makes AC-Router different from traditional MoE routers.
While methods like Top-$k$ routing \cite{shazeer2017outrageously, fedus2022switch, dai2024deepseekmoe, shi2025time} rely on a hyperparameter $k$ to strictly define the number of active experts per sequence, we argue that enforcing a hard cardinality ceiling throttles fusion diversity. Instead, we propose a soft cardinality penalty that encourages the AC-Router to select between $[1, k]$ exits organically. We estimate the number of active exits $C$ via a continuous Sigmoid surrogate, and apply an asymmetric Two-Sided Hinge Loss to penalize deviations from the desired range:
\begin{equation} \label{eq:hinge}
    \begin{aligned}
        C &= \sum_{l=1}^L \sigma(\tau \cdot \hat{\mathbf{z}}_l),\\
        \mathcal{L}_{Hinge} &= \max(0, C - k) + \gamma \max(0, 1 - C),
    \end{aligned}
\end{equation}
where $\sigma(\cdot)$ is the Sigmoid function, $\tau$ is a temperature scalar defining distribution sharpness (\eg $\tau = 10$), and $\gamma \geq 1$ controls the strength of the lower-bound penalty. This formulation ensures that if $C > k$, the first term generates gradients to reduce logit magnitudes, while if $C < 1$, the second term penalizes the network to prevent routing failure. Setting $\gamma > 1$ (\eg $\gamma = 2$) strongly discourages the zero-layer assignment problem, which is further guaranteed mechanically by the safe logits in Equation \ref{eq:safe_logits}.

Unlike recent continuous routing approaches \cite{zhuang2026ld} that rely on complex simplex projections \cite{laha2018controllable} and dedicated auxiliary MLPs to dynamically predict threshold bounds, 
our target-k hinge loss achieves cardinality regulation organically through gradient shaping. It leverages an intuitive, fully differentiable behavioral penalty to encourage exit utilization within the target range $[1,k]$, 
requiring no additional parameter overhead.

\textbf{Load Balancing Loss.} 
To prevent router collapse, where the network greedily routes traffic to a small subset of exits \cite{mu2025comprehensive, cai2025survey}, standard MoE models \cite{fedus2022switch} rely on discrete load balancing objectives. Because our AC-Router generates continuous weights without discrete sorting, we formulate a continuous analog utilizing a Softmax proxy:
\begin{equation} \label{eq:lb}
    \begin{aligned}
        \bar{P}_l &= \frac{1}{B} \sum_{b=1}^B \text{Softmax}(\mathbf{z}^{(b)})_l,\\
        \mathcal{L}_{LB} &= L \sum_{l=1}^L \bar{P}_l^2,
    \end{aligned}
\end{equation}
where $\mathbf{z}^{(b)}$ are the raw router logits for the $b$-th user sequence in a batch of size $B$. The Softmax transformation enforces a sum-to-one competition among exits, preventing an optimization loophole where the network trivially minimizes the penalty by driving all independent logits negative. Minimizing the variance of this proxy distribution $\bar{P}$ successfully encourages balanced utilization across all exits.

\textbf{Projection Anchoring (Z-Loss).} 
Mixed-precision training (\eg FP16/BF16) is a mandatory standard for scaling LLMs efficiently \cite{micikevicius2018mixed, shoeybi2019megatron, touvron2023llama}. However, during mixed-precision MoE training, unconstrained routing logits naturally diverge to extreme magnitudes, causing exponential activations to overflow and destabilize the network \cite{zoph2022st, chowdhery2023palm}. While discrete MoE models rely on Z-loss to prevent Softmax partition overflow \cite{zoph2022st}, it serves an even more critical structural role in our continuous architecture. Because our target-$k$ hinge loss (Equation \ref{eq:hinge}) relies on Sigmoid surrogates, extreme logit magnitudes cause catastrophic vanishing gradients, rendering the cardinality penalty blind. We therefore apply a simplified $L_2$ Z-loss directly to the raw routing logits:
\begin{equation} \label{eq:z_loss}
    \mathcal{L}_{Z} = \text{mean}(\mathbf{z}^2).
\end{equation}
Minimizing this loss effectively bounds pre-activation representations near zero, ensuring numerical stability, preserving steep gradients for the cardinality estimator, and preventing degenerate routing solutions.

\subsection{Model Training} \label{sec:model_train}
We visualize the framework overview in Figure \ref{fig:framework}. The model training procedure is divided into three stages:

\textbf{Stage 1: LLM Recommendation Backbone Training.}
In this stage, we perform classic E4SRec \cite{li2023e4srec} finetuning, wherein only the item ID projection layer $f$, the LoRA \cite{hu2022lora} trainable components, and the final exit $h_{L}$ are actively trained. We use the originally designed cross-entropy loss and the last exit prediction score distribution $\mathbf{O}^L$ for optimization. Note that for brevity, we omit the user index $u$ from the layer-wise score distributions (\eg $\mathbf{O}^L$ and $\mathbf{O}^l$) throughout our formulations:
\begin{equation} \label{eq:rec_loss}
\mathcal{L}_{rec} = - \sum_{u \in \mathcal{U}} \log \left( \mathbf{O}^{L}_{s_{t+1}^u} \right),
\end{equation}
where $s_{t+1}^u$ is the ground truth next-item for user $u$, $\mathbf{O}^{L}_{s_{t+1}^u}$ denotes the predicted probability assigned specifically to $s_{t+1}^u$.

\textbf{Stage 2: Intermediate Exits Training.}
Once the backbone model is adequately trained, we freeze the ID projector $f$, the LoRA modules, and the final exit $h_L$. In training stage 2, we instantiate the set of intermediate exits after each LLM layer except the final layer, denoted as $\mathcal{H}_{inter} = \mathcal{H} \setminus \{ h_L\}$. The objective of this stage is to force each intermediate exit to independently learn how to map its corresponding layer's contextualized embedding, $\mathbf{E}^l_{[-1]}$, into a valid item ranking distribution. The forward pass is executed layer-by-layer, and the logits extracted after each layer are passed through a softmax function to obtain the layer-wise distributions $\mathbf{O}^{l}$. The loss function is defined as the mean cross-entropy across all $L-1$ intermediate exits:
\begin{equation}
\mathcal{L}_{inter} = \frac{1}{L-1} \sum_{l = 1}^{L- 1} \sum_{u \in \mathcal{U}} - \log \left( \mathbf{O}^{l}_{s_{t+1}^u} \right),
\end{equation}
where $\mathbf{O}^{l}_{s_{t+1}^u}$ denotes the predicted probability assigned to the ground-truth next item $s_{t+1}^u$ by the $l$-th intermediate exit.
The strategy to train the single-exit recommender first, followed by the co-training of the intermediate exits, is inspired by \cite{xin2020deebert}. This ensures the backbone is sufficiently trained and stabilized before warming up the intermediate layers. Consequently, we ensure that each exit point produces a mathematically stable and semantically meaningful score distribution, providing a high-quality, pre-conditioned expert pool for the router.

\textbf{Stage 3: AC-Router Training.}
In training stage 3, we freeze all LLM backbone parameters, the ID projector $f$, and all exits $\mathcal{H}$. The only trainable component is the AC-Router's projection matrix $\mathbf{W}_z$. We optimize the three router training objective functions introduced in Section \ref{sec:ACR_learn}:
\begin{equation} 
    \mathcal{L}_{router} = \alpha \mathcal{L}_{LB} + \lambda \mathcal{L}_{Hinge} + \beta \mathcal{L}_{Z},
\end{equation}
where $\alpha, \lambda, \beta$ control the penalty strengths of load balance, target-$k$ hinge and projection anchoring respectively; 
alongside the recommendation loss $\mathcal{L}_{rec}$ defined in Equation \ref{eq:rec_loss}, with the score distribution changed to $\mathbf{O}^{fused}$. Because the exits are already competent rankers, the router's gradients are purely a reflection of its fusion strategy. This allows the AC-Router to efficiently learn optimal, sequence-specific exit combinations without interference from shifting score distributions. 
Additionally, because this final tuning stage is computationally lightweight, it unlocks a highly flexible deployment paradigm. One can efficiently train a suite of distinct routers, each trained with a different target-$k$ value to accommodate the specific computational constraints and performance requirements of various devices, while sharing the exact same frozen backbone obtained from the previous two training stages.

%% file: sections/experiment.tex
In this section, we document the experiments conducted to verify the performance and efficiency merits of our proposed method FLEXRec. We organize this section to answer four research questions (RQs): 
\begin{itemize}
    \item \textbf{RQ1:} Does FLEXRec outperform traditional and other LLM-based recommender systems?
    \item \textbf{RQ2:} What is the effectiveness of the proposed components in FLEXRec?
    \item \textbf{RQ3:} How sensitive is FLEXRec to different hyperparameters?
    \item \textbf{RQ4:} How does FLEXRec fuse the exits for various user sequences?
\end{itemize}

\subsection{Experiment Settings} \label{sec:exp_setting}
\input{tables/dataset}
\textbf{Datasets.} Experiments are carried out on three public benchmark datasets: Amazon review data \cite{he2016ups} --- Toys and Games (Toys) and Beauty, as well as the Yelp review dataset \footnote{https://business.yelp.com/data/resources/open-dataset/}. The statistics of the datasets are shown in Table 
\ref{tab:dataset_stats}. We follow the preprocessing practices described in \cite{li2023e4srec, zhou2020s3} to apply the 5-core item filtering. For data partition, we follow \cite{kang2018self, li2023e4srec} to perform the leave-one-out strategy, wherein for each user sequence, the last and second-last items are treated as test and validation data and the rest are used in training.

\textbf{Evaluation Metrics.} To evaluate the recommendation performance, we adopt the commonly used metrics \textbf{NDCG@N} and \textbf{Recall@N} with $N$ set to $\{10, 20\}$. We also record the average user inference time for each method to compare the inference efficiency.

\textbf{Baseline Methods.}
We evaluate FLEXRec against traditional sequential models (\textbf{SASRec} \cite{kang2018self}, \textbf{BERT4Rec} \cite{sun2019bert4rec}) and state-of-the-art LLM-based recommenders. The latter includes \textbf{E4SRec} \cite{li2023e4srec} (our static, single-exit counterpart), \textbf{SPRec} \cite{gao2025sprec} (self-play debiasing), \textbf{SETRec} \cite{lin2025order} (order-agnostic modeling), \textbf{LLM-SRec} \cite{kim2025lost} (knowledge distillation), and \textbf{HUM} \cite{bao2025heterogeneous} (heterogeneous user modeling). In addition, to isolate our AC-Router's efficacy, we substitute it with alternative MoE routing strategies implemented identically on our multi-layer architecture: discrete Top-$k$ routing (\textbf{SparseToken} \cite{shazeer2017outrageously}), threshold-based dynamic routing (\textbf{DynamicMoE} \cite{huang2024harder}), and simplex-projected continuous routing (\textbf{LD-MoLE} \cite{zhuang2026ld}). All LLM frameworks are tested on small-scale backbones (Qwen 3 1.7B \cite{yang2025qwen3}, Llama 3.2 3B \cite{grattafiori2024llama}). We also evaluate vanilla E4SRec with larger backbones (\ie Qwen 3 4B, Llama 3.1 8B) to serve as a \textbf{Skyline} performance indicator.

\textbf{Implementation Details.} For FLEXRec, we initialize item ID embeddings using a pretrained SASRec model with a dimension size of $64$, which are then linearly projected to match the hidden size of the LLM backbone. In Stage 1, we perform LoRA \cite{hu2022lora} finetuning on the LLM backbone for $2$ epochs. In Stage 2, we train the intermediate exits for a maximum of $5$ epochs, selecting the checkpoint that yields the best average performance across all layers. In Stage 3, we train the AC-Router for $2$ epochs, implementing an early stopping patience of $10$ evaluations, with model validation performed every $500$ steps. Across all stages, the global batch size is fixed at $128$, using a micro-batch size of $32$ for the Toys and Yelp datasets, and $64$ for the Beauty dataset. We utilize the AdamW optimizer with a cosine learning rate scheduler and an initial learning rate of $\expnumber{3}{-4}$. For FLEXRec's routing hyperparameters, we set the target $k = 3$ and fix the structural constants as $\xi = \expnumber{1}{-9}$, $\epsilon = 0.1$, $\tau = 10$, and $\gamma = 2$. We fine-tune the loss weights $\alpha \in \{0.005, 0.01, 0.05, 0.1, 0.15\}$, $\beta \in \{\expnumber{1}{-5}, \expnumber{1}{-4}, \expnumber{1}{-3}, \expnumber{1}{-2}, \expnumber{1}{-1}\}$, and $\lambda \in \{0.5, 1, 1.5, 2, 2.5\}$. All experiments are conducted on a single NVIDIA H100 PCIe (80GB) GPU.

\subsection{Overall Performance (RQ1)}
\input{tables/overall}

\begin{figure}[t]
    \centering
    \begin{subfigure}{0.48\columnwidth}
        \includegraphics[width=\textwidth]{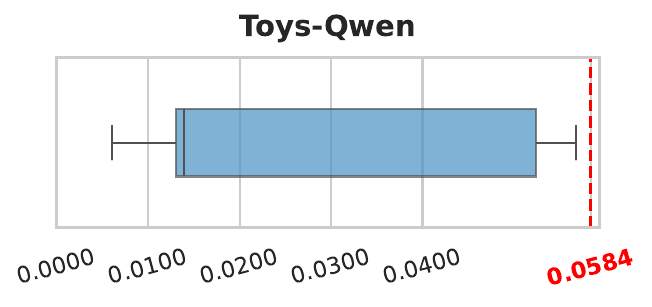}
    \end{subfigure}\hfill
    \begin{subfigure}{0.48\columnwidth}
        \includegraphics[width=\textwidth]{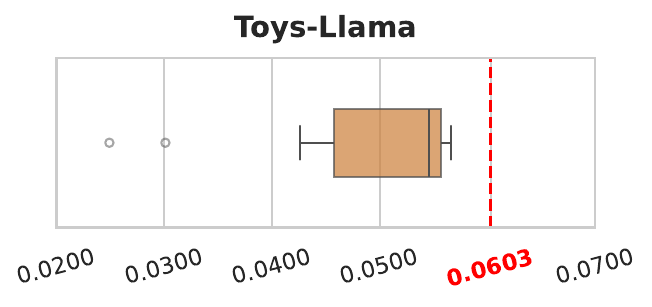}
    \end{subfigure}

    \vspace{2mm} 
    
    \begin{subfigure}{0.48\columnwidth}
        \includegraphics[width=\textwidth]{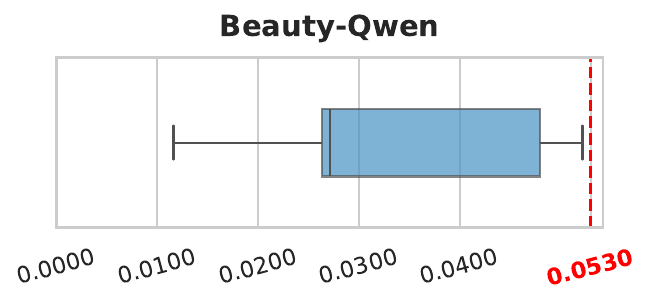}
    \end{subfigure}\hfill
    \begin{subfigure}{0.48\columnwidth}
        \includegraphics[width=\textwidth]{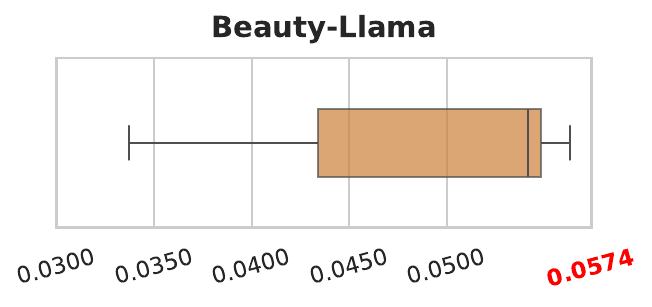}
    \end{subfigure}

    \vspace{2mm} 
    
    \begin{subfigure}{0.48\columnwidth}
        \includegraphics[width=\textwidth]{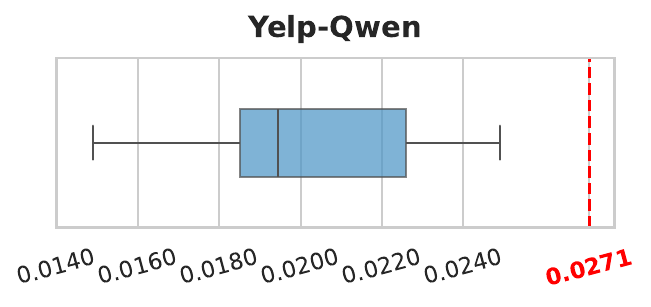}
    \end{subfigure}\hfill
    \begin{subfigure}{0.48\columnwidth}
        \includegraphics[width=\textwidth]{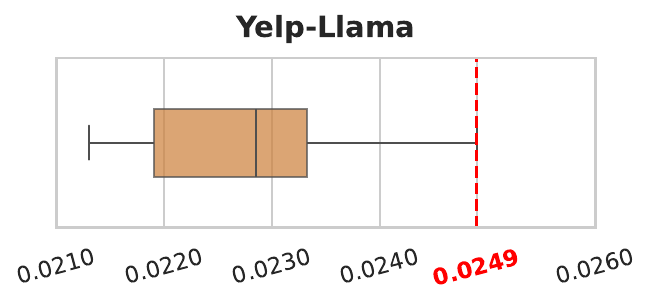}
    \end{subfigure}
    
    \caption{Performance distribution (NDCG@20) of all independent exits across datasets and backbones. The red dashed line, with its exact magnitude marked in bold red on the $x$-axis, denotes the performance of FLEXRec.}
    \label{fig:rq1_boxplots}
    \vspace*{-3mm}
\end{figure}
We present the overall performance results of our work and baseline methods in Table \ref{tab:rq1_performance}. First, we observe that the Skyline setting consistently outperforms the standard E4SRec baseline, confirming that the parameter scale of the LLM backbone is crucial to the performance of general LLM-RS architectures. However, FLEXRec effectively subverts this reliance on sheer parameter size. 
Despite using a backbone approximately half the size of the Skyline setting, FLEXRec consistently achieves the best performance among the comparable compact-backbone methods on the $N = 10$ metrics. It additionally surpasses the larger-backbone Skyline in several settings, including Beauty under both backbones and Yelp under Qwen.
To verify the effectiveness of layer-wise exit fusion instead of simply selecting the best exits after the Stage 1 and 2 training, we also present the distribution plots of all independent exits in Figure \ref{fig:rq1_boxplots}.
Notably, FLEXRec’s adaptive fusion consistently outperforms the best-performing individual exit across all evaluated settings.

When comparing our framework against MoE variants (\ie SparseToken, DynamicMoE, and LD-MoLE), we observe that while the variants generally improve upon the original E4SRec on Toys and Beauty, they falter on the sparser Yelp dataset, especially for SparseToken and DynamicMoE variants. As for other LLM methods, their performance varies greatly and generally falls short of discriminative models like E4SRec and FLEXRec. Some methods, such as SPRec, even underperform traditional models.
One possible explanation is their greater sensitivity to item textual representations, which may affect sample selection during DPO training.

Regarding model inference efficiency, traditional methods naturally incur the lowest latency due to their architectural simplicity. Among LLM-based approaches, the discriminative E4SRec remains the fastest. This showcases the major merit of discriminative models compared to generative ones (\eg SETRec), which rely on time-consuming autoregressive text decoding. Our method inherits this efficiency paradigm. By processing the sequence in a single forward pass and utilizing lightweight routing, FLEXRec introduces minimal additional latency (less than 1ms overhead compared to E4SRec) while delivering impressive accuracy gains.

\subsection{Ablation Study (RQ2)}
\input{tables/ablation}

\begin{figure*}[t]
    \centering
    \begin{minipage}{.3\textwidth}
        \centering
        \includegraphics[width=\textwidth]{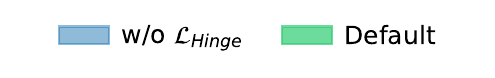}
        \vspace*{-8mm}
    \end{minipage}

    \begin{subfigure}{0.15\textwidth}
        \includegraphics[width=\textwidth]{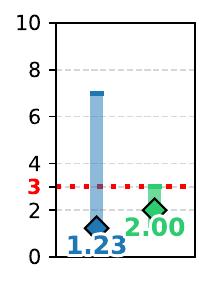}
        \vspace*{-8mm}
        \caption[]{{\small Toys-Qwen}}
    \end{subfigure}\hfill
    \begin{subfigure}{0.15\textwidth}
        \includegraphics[width=\textwidth]{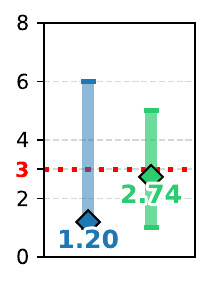}
        \vspace*{-8mm}
        \caption[]{{\small Toys-Llama}}
    \end{subfigure}\hfill
    \begin{subfigure}{0.15\textwidth}
        \includegraphics[width=\textwidth]{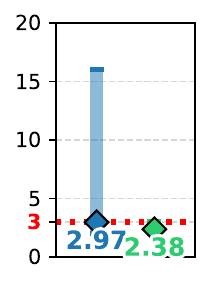}
        \vspace*{-8mm}
        \caption[]{{\small Beauty-Qwen}}
    \end{subfigure}\hfill
    \begin{subfigure}{0.15\textwidth}
        \includegraphics[width=\textwidth]{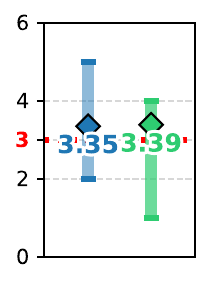}
        \vspace*{-8mm}
        \caption[]{{\small Beauty-Llama}}
    \end{subfigure}\hfill
    \begin{subfigure}{0.15\textwidth}
        \includegraphics[width=\textwidth]{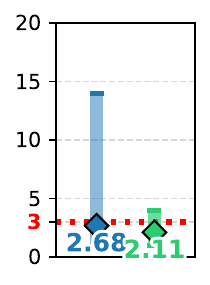}
        \vspace*{-8mm}
        \caption[]{{\small Yelp-Qwen}}
    \end{subfigure}\hfill
    \begin{subfigure}{0.15\textwidth}
        \includegraphics[width=\textwidth]{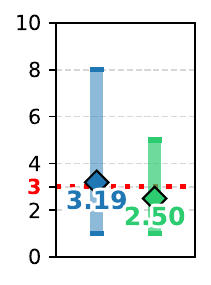}
        \vspace*{-8mm}
        \caption[]{{\small Yelp-Llama}}
    \end{subfigure}
    
    \caption{Impact of the target-$k$ hinge loss ($\mathcal{L}_{Hinge}$) on the number of active exits across inference samples. The y-axis denotes the number of active exits assigned by the AC-Router for a given user sequence. The vertical bars represent the minimum to maximum range of exits assigned across all samples, while the diamond markers denote the average number of exits used. The red dotted horizontal line represents the target number of exits ($k=3$).}
    \label{fig:rq2_hinge_loss}
\end{figure*}

We conduct ablation studies on the three AC-Router-related loss terms in Section \ref{sec:ACR_learn} to demonstrate their importance in training an optimal router for layer-wise exit fusion. Table \ref{tab:ablation_study} presents the performance of the default framework alongside variants where individual loss components are disabled. The empirical results confirm that the default setting consistently achieves the highest accuracy across all datasets and both LLM backbones. Disabling any of the introduced loss terms leads to a sub-optimally trained router and inevitably impedes overall recommendation performance.

Additionally, we investigate the impact of the proposed target-$k$ hinge loss ($\mathcal{L}_{Hinge}$) in controlling the number of exits utilized for fusion. Figure \ref{fig:rq2_hinge_loss} plots the exit usage statistics to compare the default setting against the variant lacking $\mathcal{L}_{Hinge}$. Removing this hinge loss clearly causes severe routing instability. For instance, the maximum number of active exits surges to 16 on the Beauty dataset under the Qwen backbone. 
In contrast, the default configuration substantially reduces extreme exit activation and keeps the average utilization near the target capacity.

\subsection{Hyperparameter Study (RQ3)}
\begin{figure*}[t]
    \centering
    \begin{minipage}{.4\textwidth}
        \centering
        \includegraphics[width=.9\textwidth]{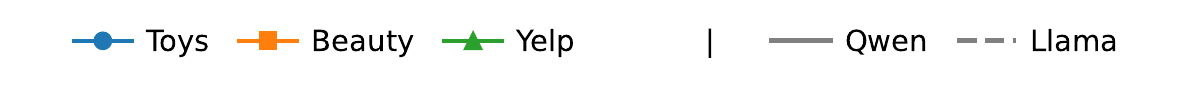}
        \vspace*{-6mm}
    \end{minipage}

    \includegraphics[width=\textwidth]{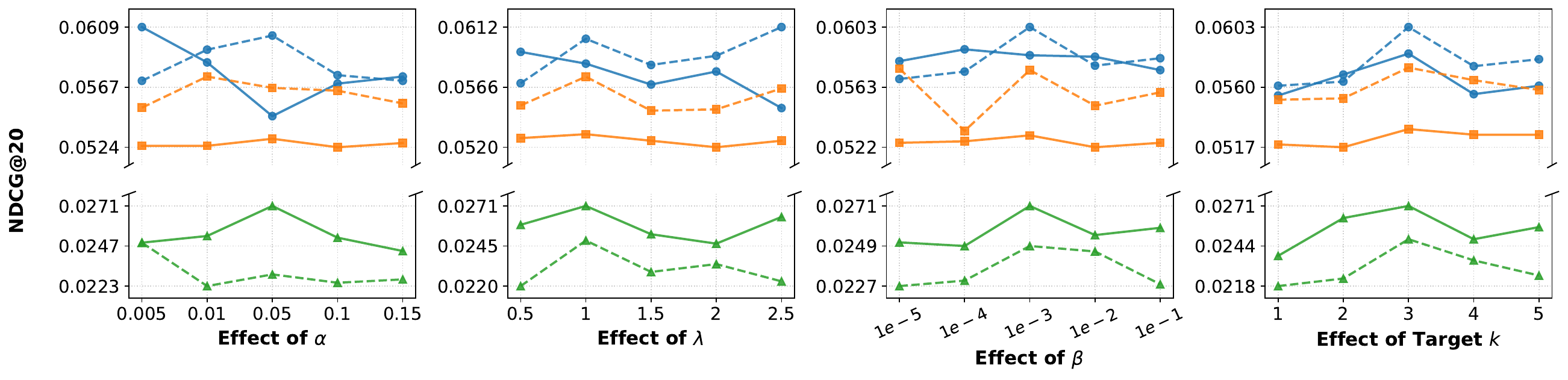}
    \caption{Impact of different hyperparameters ($\alpha$, $\lambda$, $\beta$, and $k$) on the recommendation performance (NDCG@20) of FLEXRec. 
    Solid and dashed lines represent the Qwen and Llama backbones, respectively. A split-axis design dynamically scales the y-axis to visually separate the higher-performing datasets (Toys and Beauty) from the sparser dataset (Yelp).
    }
    \label{fig:hyperparameters}
    \vspace*{-2mm}
\end{figure*}

We investigate FLEXRec's sensitivity to various hyperparameter configurations. Specifically, we evaluate the impact of the load balancing loss weight ($\alpha$), the target-$k$ hinge loss weight ($\lambda$), the Z-loss weight ($\beta$), and the target number of exits ($k$). We tune these hyperparameters within the following ranges: $\alpha \in \{0.005, 0.01, 0.05, 0.1, 0.15\}$, $\lambda \in \{0.5, 1, 1.5, 2, 2.5\}$, $\beta \in \{\expnumber{1}{-5}, \expnumber{1}{-4}, \expnumber{1}{-3}, \expnumber{1}{-2}, \expnumber{1}{-1} \}$, and $k \in \{1, 2, 3, 4, 5\}$. The overall performance trends are illustrated in Figure \ref{fig:hyperparameters}.

\textbf{Effect of the load balancing loss weight ($\alpha$).} The parameter $\alpha$ plays a critical role in FLEXRec's performance. Under the Qwen backbone, the Yelp dataset peaks at $\alpha=0.05$, whereas the Toys dataset experiences a noticeable performance dip at this value before recovering. Conversely, with the Llama backbone, $\alpha=0.05$ yields the optimal result for Toys. These divergent trends suggest that the optimal $\alpha$ is highly dependent on the interplay between the dataset's characteristics and the underlying LLM architecture.

\textbf{Effect of the target-$k$ hinge loss weight ($\lambda$).} The empirical results demonstrate a general preference for $\lambda=1$, at which all three datasets achieve robust recommendation accuracy. As $\lambda$ increases beyond this optimal point, we observe a marginal performance degradation, which is particularly evident on the Toys dataset under the Qwen backbone and the Yelp dataset under the Llama backbone.

\textbf{Effect of the Z-loss weight ($\beta$).} Setting $\beta = \expnumber{1}{-3}$ consistently delivers optimal or near-optimal results across all evaluated scenarios. The stabilizing effect of this regularizer is most prominent on the Beauty dataset under the Llama backbone: the performance suffers a sharp decline at $\beta = \expnumber{1}{-4}$ but immediately recovers and peaks when $\beta$ is increased to $\expnumber{1}{-3}$.

\textbf{Effect of the target number of exits ($k$).} Across all datasets and both LLM backbones, setting target $k=3$ consistently yields the highest recommendation accuracy. This indicates that dynamically routing and fusing roughly three exits strikes the optimal balance between extracting sufficient semantic richness and preventing the introduction of noisy, redundant signals into the learning process.

\begin{figure}[!b]
    \centering
    \includegraphics[width=\columnwidth, keepaspectratio]{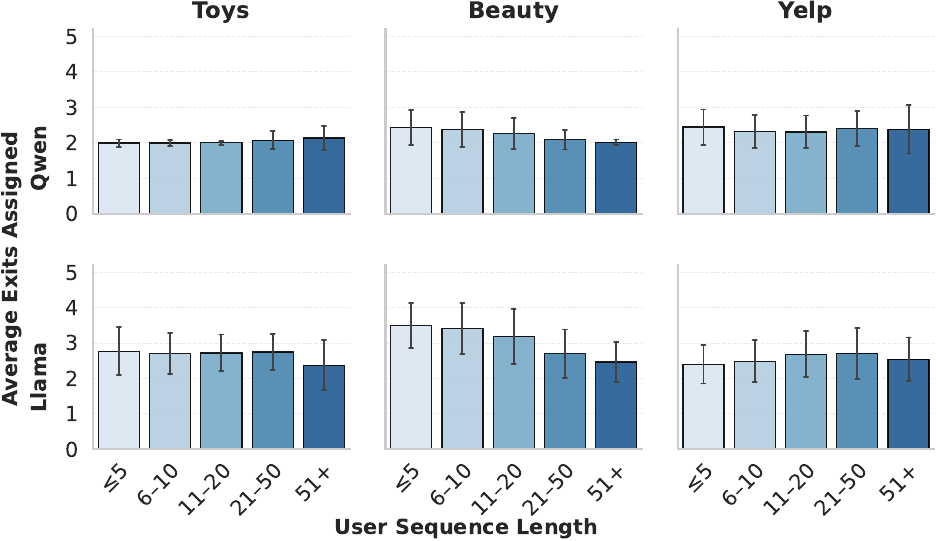}
    \caption{Adaptive routing behavior of FLEXRec across varying user sequence lengths. The x-axis categorizes users into interaction history bins, while the y-axis indicates the average number of exits assigned. Error bars denote the standard deviation of the router's decisions.}
    \label{fig:rq4_routing_depth}
    \vspace*{-3mm}
\end{figure}

\subsection{Adaptive Routing Case Study (RQ4)}
To study how FLEXRec assigns exits across varying user profiles, we categorize user sequences into five length-based groups and visualize the average exits assigned in Figure \ref{fig:rq4_routing_depth}.

It is observed that the number of exits used for fusion is heavily influenced by the choice of LLM backbone. 
FLEXRec consistently activates more exits when utilizing Llama (approximately 3 on average) than when utilizing Qwen (approximately 2).
This indicates that the AC-Router effectively calibrates to the inherent representational capacity and layer dynamics of the underlying LLM backbone model.

Furthermore, the standard deviation within each sequence-length group remains substantial, indicating that FLEXRec does not rely solely on rigid length-based heuristics. Instead, sequences of similar lengths can receive different routing decisions, suggesting that the router responds to sequence-specific information beyond interaction-history length. This behavior is particularly evident with the Llama backbone, where within-group variation is comparatively large.

%% file: tables/dataset.tex
\begin{table}[t]
\centering
\caption{Statistics of the datasets.}
\label{tab:dataset_stats}
\resizebox{0.48\textwidth}{!}{
\begin{tabular}{lcccc}
\toprule
\textbf{Dataset} & \textbf{\# Users} & \textbf{\# Items} & \textbf{\# Actions} & \textbf{Sparsity} \\
\midrule
Toys   & 19,412 & 11,924 & 167,597 & 99.93\% \\
Beauty & 22,363 & 12,101 & 198,502 & 99.93\% \\
Yelp   & 30,431 & 20,033 & 316,354 & 99.95\% \\
\bottomrule
\end{tabular}
}
\vspace*{-3mm}
\end{table}

%% file: tables/overall.tex
\begin{table*}[t]
\centering
\caption{The overall performance benchmark of FLEXRec and baseline methods. Skyline settings represent E4SRec evaluated with a larger LLM backbone (\ie Qwen 3 4B and Llama 3.1 8B) to establish an upper-bound performance reference. Boldface and underline indicate the best and second-best results, respectively, among the comparable methods under each backbone setting.}
\label{tab:rq1_performance}
\resizebox{\textwidth}{!}{
\begin{tabular}{ll ccccc ccccc ccccc}
\toprule
\multirow{2}{*}{\textbf{LLM}} & \multirow{2}{*}{\textbf{Method}} & \multicolumn{5}{c}{\textbf{Toys}} & \multicolumn{5}{c}{\textbf{Beauty}} & \multicolumn{5}{c}{\textbf{Yelp}} \\
\cmidrule(lr){3-7} \cmidrule(lr){8-12} \cmidrule(lr){13-17}
 & & N@10 & R@10 & N@20 & R@20 & T(ms) & N@10 & R@10 & N@20 & R@20 & T(ms) & N@10 & R@10 & N@20 & R@20 & T(ms) \\
\midrule
\multirow{2}{*}{\NA} & SASRec & 0.0353 & 0.0649 & 0.0421 & 0.0919 & 0.01 & 0.0288 & 0.0571 & 0.0376 & 0.0922 & 0.01 & 0.0173 & 0.0354 & 0.0233 & 0.0593 & 0.01 \\
 & BERT4Rec & 0.0210 & 0.0385 & 0.0257 & 0.0571 & 0.01 & 0.0228 & 0.0458 & 0.0287 & 0.0695 & 0.01 & 0.0174 & 0.0345 & 0.0232 & 0.0578 & 0.01 \\
\midrule
\multirow{10}{*}{Qwen} & Skyline & 0.0539 & 0.0867 & 0.0616 & 0.1172 & 9.51 & 0.0435 & 0.0745 & 0.0510 & 0.1045 & 9.49 & 0.0199 & 0.0400 & 0.0260 & 0.0643 & 9.83\\
\cmidrule{2-17}
 & SPRec & 0.0170 & 0.0328 & 0.0208 & 0.0478 & 238.05 & 0.0073 & 0.0140 & 0.0092 & 0.0217 & 244.36 & 0.0099 & 0.0109 & 0.0102 & 0.0125 & 77.65 \\
 & SETRec & 0.0423 & 0.0760 & 0.0510 & 0.1107 & 130.94 & 0.0380 & 0.0714 & 0.0462 & 0.1037 & 141.49 & 0.0177 & 0.0358 & 0.0239 & 0.0605 & 139.81 \\
 & LLM-SRec & 0.0255 & 0.0485 & 0.0322 & 0.0753 & 62.01 & 0.0198 & 0.0412 & 0.0263 & 0.0674 & 67.54 & 0.0148 & 0.0307 & 0.0203 & 0.0529 & 46.10 \\
 & HUM & 0.0269 & 0.0565 & 0.0349 & 0.0883 & 23.43 & 0.0204 & 0.0444 & 0.0280 & 0.0746 & 24.38 & 0.0136 & 0.0249 & 0.0172 & 0.0391 & 23.91 \\
 & E4SRec & 0.0452 & 0.0742 & 0.0515 & 0.0990 & 6.89 & 0.0391 & 0.0694 & 0.0470 & 0.1006 & 7.00 & 0.0188 & \underline{0.0383} & 0.0248 & \underline{0.0621} & 6.98 \\
\cmidrule{2-17}
 & SparseToken & 0.0499 & 0.0803 & 0.0561 & 0.1050 & 7.31 & \underline{0.0445} & \underline{0.0763} & \underline{0.0526} & \underline{0.1083} & 7.40 & 0.0180 & 0.0358 & 0.0239 & 0.0594 & 7.69 \\
 & DynamicMoE & 0.0497 & 0.0791 & 0.0560 & 0.1040 & 7.35 & 0.0426 & 0.0747 & 0.0506 & 0.1063 & 7.37 & \underline{0.0191} & 0.0381 & \underline{0.0251} & 0.0619 & 7.87 \\
 & LD-MoLE & \underline{0.0516} & \underline{0.0816} & \underline{0.0578} & \underline{0.1064} & 7.60 & 0.0438 & 0.0753 & 0.0513 & 0.1054 & 6.63 & 0.0184 & 0.0365 & 0.0242 & 0.0598 & 6.68 \\
\cmidrule{2-17}
 & \textbf{FLEXRec} & \textbf{0.0521} & \textbf{0.0830} & \textbf{0.0584} & \textbf{0.1084} & 7.53 & \textbf{0.0451} & \textbf{0.0775} & \textbf{0.0530} & \textbf{0.1091} & 7.10 & \textbf{0.0207} & \textbf{0.0407} & \textbf{0.0271} & \textbf{0.0666} & 7.49 \\
\midrule
\multirow{10}{*}{Llama} & Skyline & 0.0526 & 0.0836 & 0.0595 & 0.1113 & 10.66 & 0.0471 & 0.0788 & 0.0556 & 0.1126 & 10.19 & 0.0206 & 0.0407 & 0.0268 & 0.0655 & 10.84 \\
\cmidrule{2-17}
 & SPRec & 0.0184 & 0.0331 & 0.0224 & 0.0489 & 65.86 & 0.0158 & 0.0268 & 0.0187 & 0.0383 & 144.54 & 0.0136 & 0.0156 & 0.0140 & 0.0172 & 200.84 \\
 & SETRec & 0.0446 & 0.0811 & 0.0532 & \underline{0.1153} & 126.58 & 0.0418 & 0.0768 & 0.0504 & \underline{0.1111} & 127.39 & \textbf{0.0189} & \textbf{0.0376} & \textbf{0.0250} & \underline{0.0620} & 128.84 \\
 & LLM-SRec & 0.0194 & 0.0376 & 0.0257 & 0.0628 & 53.62 & 0.0189 & 0.0384 & 0.0255 & 0.0646 & 55.44 & 0.0138 & 0.0284 & 0.0190 & 0.0493 & 37.22 \\
 & HUM & 0.0341 & 0.0724 & 0.0443 & 0.1127 & 20.58 & 0.0299 & 0.0632 & 0.0394 & 0.1011 & 21.77 & 0.0185 & 0.0339 & 0.0232 & 0.0524 & 21.02 \\
 & E4SRec & 0.0481 & 0.0773 & 0.0546 & 0.1035 & 6.48 & 0.0470 & 0.0771 & 0.0549 & 0.1084 & 6.57 & 0.0182 & 0.0366 & 0.0243 & 0.0608 & 6.78 \\
\cmidrule{2-17}
 & SparseToken & \underline{0.0521} & \underline{0.0828} & \underline{0.0592} & 0.1110 & 7.31 & \underline{0.0488} & \underline{0.0803} & 0.0563 & 0.1104 & 8.42 & 0.0181 & \underline{0.0370} & 0.0241 & 0.0611 & 7.55 \\
 & DynamicMoE & 0.0492 & 0.0793 & 0.0557 & 0.1054 & 7.36 & 0.0476 & 0.0771 & 0.0548 & 0.1055 & 7.87 & 0.0167 & 0.0340 & 0.0227 & 0.0577 & 7.07 \\
 & LD-MoLE & 0.0512 & 0.0820 & 0.0581 & 0.1095 & 7.38 & \underline{0.0488} & 0.0795 & \underline{0.0564} & 0.1094 & 7.51 & 0.0182 & 0.0369 & 0.0247 & \textbf{0.0627} & 7.14 \\
\cmidrule{2-17}
 & \textbf{FLEXRec} & \textbf{0.0524} & \textbf{0.0848} & \textbf{0.0603} & \textbf{0.1166} & 7.43 & \textbf{0.0498} & \textbf{0.0818} & \textbf{0.0574} & \textbf{0.1122} & 8.12 & \underline{0.0188} & \textbf{0.0376} & \underline{0.0249} & 0.0617 & 7.31 \\
\bottomrule
\end{tabular}
}
\end{table*}

%% file: tables/ablation.tex
\begin{table*}[t]
\centering
\caption{Ablation study of FLEXRec on three datasets using both Qwen and Llama backbones. ``Default" setting refers to the model utilizing all loss components. We evaluate the impact of removing the load balancing loss (w/o $\mathcal{L}_{LB}$), the target-$k$ hinge loss (w/o $\mathcal{L}_{Hinge}$), and the Z-loss (w/o $\mathcal{L}_Z$).}
\label{tab:ablation_study}
\resizebox{\textwidth}{!}{
\begin{tabular}{ll cccc cccc cccc}
\toprule
\multirow{2}{*}{\textbf{LLM}} & \multirow{2}{*}{\textbf{Variant}} & \multicolumn{4}{c}{\textbf{Toys}} & \multicolumn{4}{c}{\textbf{Beauty}} & \multicolumn{4}{c}{\textbf{Yelp}} \\
\cmidrule(lr){3-6} \cmidrule(lr){7-10} \cmidrule(lr){11-14}
 & & N@10 & R@10 & N@20 & R@20 & N@10 & R@10 & N@20 & R@20 & N@10 & R@10 & N@20 & R@20 \\
\midrule
\multirow{4}{*}{Qwen} & \textbf{Default} & 0.0521 & 0.0830 & 0.0584 & 0.1084 & 0.0451 & 0.0775 & 0.0530 & 0.1091 & 0.0207 & 0.0407 & 0.0271 & 0.0666 \\
 & w/o $\mathcal{L}_{LB}$ & 0.0476 & 0.0772 & 0.0540 & 0.1027 & 0.0447 & 0.0766 & 0.0528 & 0.1088 & 0.0177 & 0.0350 & 0.0235 & 0.0585 \\
 & w/o $\mathcal{L}_{Hinge}$ & 0.0504 & 0.0787 & 0.0564 & 0.1026 & 0.0444 & 0.0766 & 0.0523 & 0.1082 & 0.0190 & 0.0373 & 0.0249 & 0.0609 \\
 & w/o $\mathcal{L}_Z$ & 0.0471 & 0.0769 & 0.0536 & 0.1024 & 0.0445 & 0.0771 & 0.0523 & 0.1079 & 0.0183 & 0.0367 & 0.0241 & 0.0597 \\
\midrule
\multirow{4}{*}{Llama} & \textbf{Default} & 0.0524 & 0.0848 & 0.0603 & 0.1166 & 0.0498 & 0.0818 & 0.0574 & 0.1122 & 0.0188 & 0.0376 & 0.0249 & 0.0617 \\
 & w/o $\mathcal{L}_{LB}$ & 0.0486 & 0.0778 & 0.0554 & 0.1049 & 0.0483 & 0.0784 & 0.0559 & 0.1087 & 0.0170 & 0.0348 & 0.0229 & 0.0585 \\
 & w/o $\mathcal{L}_{Hinge}$ & 0.0488 & 0.0781 & 0.0553 & 0.1039 & 0.0485 & 0.0793 & 0.0561 & 0.1092 & 0.0166 & 0.0344 & 0.0223 & 0.0569 \\
 & w/o $\mathcal{L}_Z$ & 0.0513 & 0.0846 & 0.0592 & 0.1158 & 0.0495 & 0.0803 & 0.0567 & 0.1090 & 0.0169 & 0.0342 & 0.0225 & 0.0564 \\
\bottomrule
\end{tabular}
}
\vspace*{-2mm}
\end{table*}

%% file: sections/conclusion.tex
In this paper, we propose FLEXRec, a layer-wise exit fusion framework that enhances compact LLMs for scalable sequential recommendation. FLEXRec adopts a discriminative retrieval paradigm to enable efficient full-corpus ranking and employs an adaptive continuous router (AC-Router) to dynamically select and fuse layer-wise score distributions for each user sequence. A novel target-k hinge loss further regulates the number of activated exits to maintain routing sparsity. Extensive experiments on three real-world datasets demonstrate that FLEXRec effectively compensates for the limited capacity of compact LLM backbones, achieving state-of-the-art accuracy among compact-backbone methods while maintaining efficient inference.